\documentstyle[12pt]{article} 

\begin{document}
{\pagestyle{empty}

{\Large\begin{center} {\bf Consistency test of the Standard Model} 
\end{center}} \vskip .5cm

\begin{center} {\bf Marek Paw\l owski}$^{1\dagger}$ \\
{\bf and } \\ ~ 
\\{\bf
\frame{Ryszard R\c aczka}}\\ ~ \\
Soltan
Institute for Nuclear Studies, Warsaw, POLAND \\  ~


\end{center}

}

\vskip 2cm

\bigskip\bigskip \centerline{\bf Abstract}

\bigskip If the "Higgs mass" is not the physical mass of a real particle but
rather an effective ultraviolet cutoff then a process energy dependence of this
cutoff must be admitted.  Precision data from at least two energy scale
experimental points are necessary to test this hypothesis.  The first set of
precision data is provided by the $Z$-boson peak experiments.  We argue that the
second set can be given by 10-20GeV $e^+e^-$ colliders.  We put attention to the
special role of tau polarization experiments that can be sensitive to the "Higgs
mass" for a sample of $\sim10^8$ produced tau pairs.  We argue that such a study
may be regarded as a negative selfconsistency test of the Standard Model and of
most of its extensions.

\vfil \vskip.5cm \hrule\smallskip

{\footnotesize

\parindent 0pt

$^1$~~Supported by Polish
Committee for Scientific Researches.

$^{\dagger}$~~e--mail:  PAWLOWSK@fuw.edu.pl}

\eject

The introduction of the physical Higgs particle with nonzero mass regularizes
some ultraviolet divergences of electroweak theories.  It makes that the
Standard Model (SM) and most of its extensions are perturbatively
renormalizable.  The physical mass of the Higgs particle $m_H$ can be indirectly
predicted by the theory:  it can be derived from the precision measurements of
observables that are sensitive to the radiative corrections dependent on $m_H$.
Here $m_H$ plays in fact a role of a UV regulator for bosonic contributions.
The Higgs mass is a measurable physical constant, by definition.  Consequently
it must be the same when derived indirectly from the data of different
experiments performed at different energy scales.

The regularizing role of $m_H$ can be also played by the ultraviolet cutoff
introduced in the models where the Higgs particle is absent
\cite{I,schildknecht,herrero}.  The cutoff is an artificial element that we
introduce to cover an incompleteness of the model or the imperfection of our
calculational methods either it simply reduces a complicated task to a less
complicated one.  If we try to hide our ignorance in a simplest way we introduce
one additional parameter - a cutoff $\Lambda$.  We hope that this parameter can
be the same for a class of similar phenomena.  Predictivity of a model holds if
the cutoff $\Lambda$ is process independent for a restricted interaction energy
region at least.  It would be nice to have a universal cutoff valid for all
phenomena below some energy scale but in principle it needs not to be the case.
(We know this from the experience with QED \cite{bethe}.)  Thus we have to admit
that the cutoff is energy dependent.

In practice the energy dependence of $\Lambda$ can be studied as follows:
Predictions of a regularized model can be calculated for experiments performed
in various energy regions.  These predictions would depend on the cutoff
$\Lambda$ and inversely:  the value of the appropriate cutoff derived from
experimental data measured in different energy regions can be different.  It
makes that the cutoff becomes energy dependent in a sense.

As we have already mentioned, the UV cutoff $\Lambda$ of electroweak models is
closely connected with the Higgs mass of SM but, in contrast to the demanded and
expected energy independence of $m_H$, we have to admit an energy dependence of
$\Lambda$.  This is the difference that makes a room for experimental tests and
a comparison of the Standard Model and the models that admit an energy
dependence of effective (or dynamical) UV regulator.  The independence or
dependence of the predicted Higgs mass on a set of observables that it has been
derived from can be also seen as a sensitive selfconsistency check of the SM
itself.

Thus the task is to derive the value of UV regulator ($m_H$ or $\Lambda$) from
at least two independent sets of data collected in experiments performed in
different energy regions (to be more concrete:  with different characteristic
energy-momentum transfer of weak interactions).  The close relation between
$m_H$ and the cutoff is an essential practical and logical element of the
reasoning.  It was shown \cite{schildknecht,I} that the results of an effective
electroweak theory with the cutoff $\Lambda$ can be reproduced approximately
from the SM results when we replace $m_H$ by $\Lambda$.  The differences between
the exact and the approximate results are small and are given by known
expressions.  Thus in practice, instead of studying directly the energy
dependence of the cutoff $\Lambda$ in a cut model, we can make use of the
wealthy set of SM results.  This fact is important also from the logical point
of view.  Any observed energy dependence of the regulator $m_H$ can be clearly
interpreted as a negative result of the self-check of the SM but will be
admitted by theories without the physical Higgs boson.  Of course if no energy
dependence would be detected it will mean nothing for both classes of models.
In this sense the proposed idea constitutes a kind of negative selfconsistency
test of the SM.

The above consideration makes sense only if the Higgs particle is not found
directly.  LEP experiments put the direct limit $m_H>m_{min}$ on the mass of the
physical Higgs boson.  This direct search restriction cannot be valid longer if
we admit reinterpretation of $m_H$.  In fact the indirect LEP data are less
restrictive than the direct search limit and prefer values of $m_H$ even smaller
than $m_{min}$.  Thus in our considerations we have to ignore restrictions
following from the direct Higgs search at LEP.

The problem of choice of a renolmalization scheme must be also treated
carefully.  It is convenient to use EW on-shell renormalization scheme in SM
analysis.  Then the theory is described by the parameter set consisting of
$\alpha$, $\alpha_s$, $m_Z$, $m_W$, $m_H$ and masses and mixing angles of
fermions.  Because W-meson mass $m_W$ is known with relatively big experimental
error it is practical to replace $m_W$ by the precisely measured muon decay
constant $G_\mu$.  This quantity can be calculated within the model and one
obtains the famous relation

$$ m_W^2 = {\pi\alpha \over \sqrt{2} G^\mu \sin^2 \theta_W (1-\triangle r)}
\eqno (1) $$ where $\sin^2 \theta_W = 1-m^2_W/m^2_Z$ and $\triangle r$ contains
radiative corrections depending on all parameters of the theory including $m_H$
and $m_W$.  Equation (1) can be solved iteratively giving

$$ m_W = m_W(\alpha, \alpha_s, m_Z, G_\mu, m_H, ...)  \eqno (2) $$ 
and we can replace $m_W$ by $G_\mu$ in the parameter set of the model.

The situation is slightly different when we have to do with an effective theory
without the physical Higgs particle in which $m_H$ is not a mass of physical
particle but can be a cutoff.  Consider for example a generic quantity $\Sigma$
describing an electroweak process proceeding with characteristic energy
$E_{(1)}$.

$$
\Sigma=\Sigma(\alpha, \alpha_s, m_Z, G_\mu, {m_H}_{(1)}, ...).  \eqno
(3) $$ 

Assume that in the considered model we can derive relation analogical to (2) 

$$ m_W = m_W(\alpha, \alpha_s, m_Z, G_\mu, {m_H}_{(0)}, ...).  \eqno (4) $$ 

Relation (4) follows from the analysis of $\mu$ decay within the cut model. 
The characteristic energy for this process is $\mu$
meson mass $E_{(0)}=m_\mu\approx 0$.  Thus ${m_H}_{(1)}$ in (3) and
${m_H}_{(0)}$ in (4) need not be the same as the values of cutoff in principle
can be different for the processes with energies $E_{(1)}$ and
$E_{(0)}$.  This is the difference between the SM and models without
the Higgs boson.  We have to take into account this difference when we look for
a supposed energy dependence of $m_H$ or we can avoid this problem working with
$m_W$ as an input parameter.  The last approach is also justified by growing
accuracy of $m_W$ measurements.
\medskip

Currently the most precise EW data come from $e^+e^-$ collider experiments.  LEP
and SLC provided us with a set of information about the physics near $Z_0$ peak
with accuracy sensitive to EW radiative corrections.  A limited information
about the Higgs mass can be also derived from this data.  Many other EW
experiments were performed in the past \cite{langackerbook} but none of them was
accurate enough to give even qualitative information about $m_H$.  New
generation experiments are necessary both below and far above the energy of $Z$
mass.  We are to restrict ourselves only to low energy region in the present
analysis.

Some of the quantities measured at $Z_0$ peak can be measured in principle for
the whole $e^+e^-$ energy collision range.  These are cross sections and
production asymmetries.  Unfortunately most of them are almost insensitive to
the value of the Higgs mass except the small region near the peak.  For example
it was known since the analysis of PETRA/PEP experiments that the
forward--backward asymmetry is sensitive weaker than 0.2\% for the variation of
$m_H$ from 10GeV to 1000GeV \cite{hollik84}.

The exceptions are leptonic asymmetries, especially tau polarization $A_{pol}$
and tau polarization forward--backward asymmetry ${A_{pol}}^{FB}$.

We have enumerated the energy dependence of these quantities for $m_H=10GeV$ and
$m_H=1000GeV$.  For this purpose we have used the old version 453 of ZFITTER
package \cite{zfitter} that, according to its authors \cite{bardinprivate}, can
be applied for collision energies above $b$ pair production threshold.  We have
modified this code introducing the experimental value of $m_W$ as an input
parameter instead of conventionally used $G_\mu$ for the reasons already
discussed.  This enlarges the error but in fact only qualitative results are
interesting at the present stage.  We have check that this modification has
quantitatively observable but qualitatively unimportant consequences for the
present analysis.

We have plotted our results in Figs.  1--6.

Fig.  1 (resp.  2) shows the energy dependence of $A_{pol}$ (resp.
${A_{pol}}^{FB}$) for $m_H=10GeV$ (dashed line) and $m_H=1000GeV$
(solid line) in the energy range $\sqrt{s}<100GeV$.

\begin{picture}(40000,17000) \includegraphics{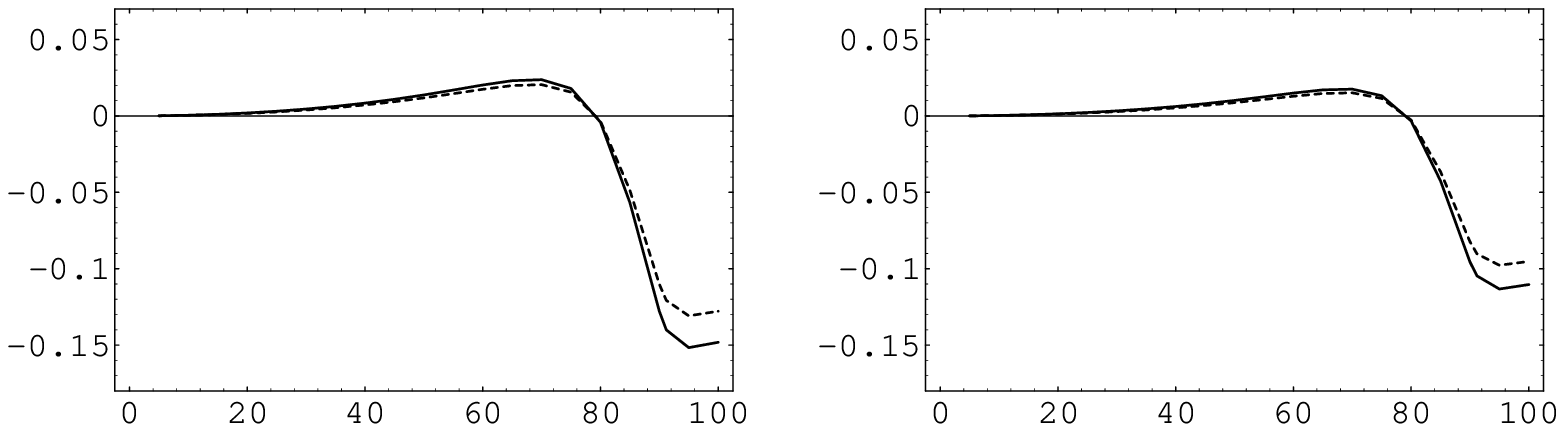} \put(35500,0000){$\sqrt{s}$}
\put(15300,0000){$\sqrt{s}$} \put(-500,14600){$A_{pol}$}
\put(20000,14600){${A_{pol}}^{FB}$}

\end{picture} \vskip-.5cm \begin{center} {\bf ~~~~Fig.  1 \hskip 6cm
Fig.  2 } \end{center} \bigskip

The same but for the restricted energy range $\sqrt{s}<60GeV$ is plotted
in Fig.  3 (resp.  4).

\begin{picture}(40000,18000) \includegraphics{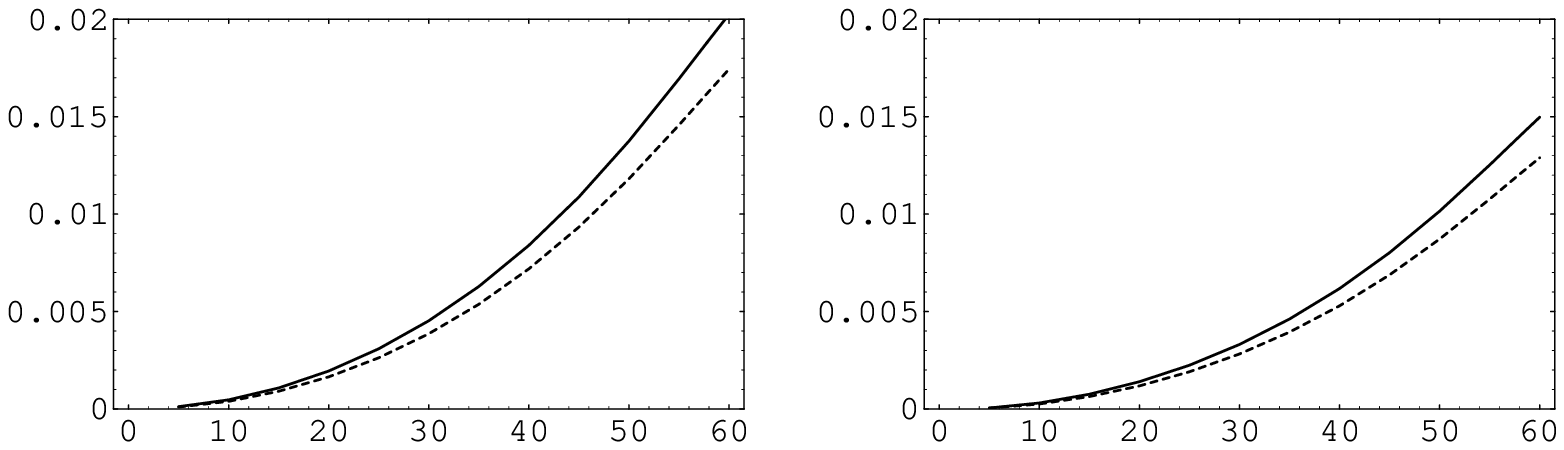} \put(35500,0000){$\sqrt{s}$}
\put(15300,0000){$\sqrt{s}$} \put(-500,14600){$A_{pol}$}
\put(20000,14600){${A_{pol}}^{FB}$}

\end{picture} \vskip-.5cm \begin{center} {\bf Fig.  3 \hskip 6cm Fig.
4 } \end{center} \bigskip

We see that the relative sensitivity to the value of the Higgs mass remains at
the same level and is bigger than $\sim$15\% although the considered quantities
rapidly decrease when the energy decreases.  This fact is shown in Fig.  5
(resp.  6) where the difference to the mean value ratio ${\cal R} A =
{A[1000]-A[10]\over {1\over 2}( A[1000]+A[10])}$ is plotted.

\begin{picture}(40000,17000) \includegraphics{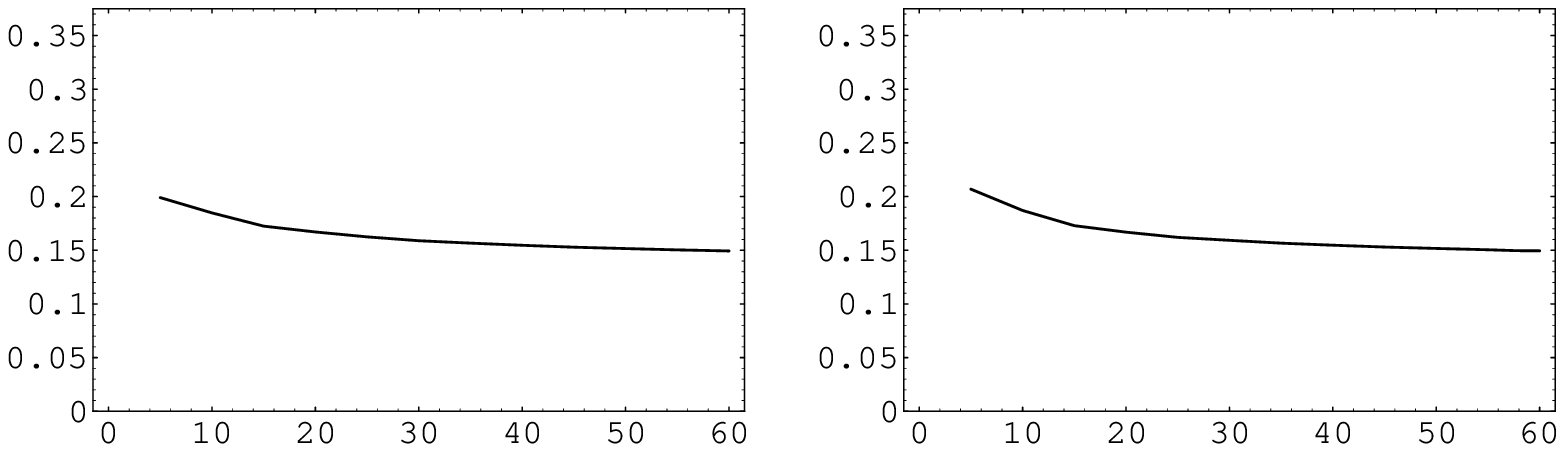} \put(35500,0000){$\sqrt{s}$}
\put(15300,0000){$\sqrt{s}$} \put(-500,14600){${\cal R} A_{pol}$}
\put(20000,14600){${\cal R}{A_{pol}}^{FB}$}

\end{picture} \vskip-.5cm \begin{center} {\bf ~~~~Fig.  5 \hskip 6cm
Fig.  6 } \end{center} \bigskip

One can try to estimate the tau pair production sample that is necessary to
observe such effect and to derive an information about $m_H$ from it.  This
estimation can be deduced from only statistical considerations.  Let quantity
$A\ll 1$ be constructed from two independently counted numbers of events $N_1$
and $N_2$ as the ratio

$$A = {N_1-N_2 \over N_1+N_2}.  \eqno (5) $$
Statistical error of $A$ may be estimated by 

$$ \triangle
A\approx {\sqrt{N}\over N} \eqno (6) $$ 
where $N=N_1+N_2$.  Taking into account the 
fact that sensitivity of $A$ to the value of $m_H$ is of order of 15\%
we demand that 

$$ \triangle A/A<0.1 \eqno (7) $$.

Both considered quantities $A_{pol}$ and ${A_{pol}}^{FB}$ are of the order or
bigger than $0.001$ for collision energies $\sqrt{s}\sim 10GeV$.  It is
easy to obtain from (6) and (7) that the event sample necessary for deriving
conclusions concerning $m_H$ from measurements of tau polarization and tau
polarization forward--backward asymmetry is of the order of $N_\tau=10^8$.  
This corresponds to the expected yearly sample at b-factory of luminosity
of the order of 10$^{34}cm^{-2}s^{-1}$. The
estimation is very rough and does not include many practical and theoretical
problems.  However it provides an information about required technical
conditions that must be fulfilled to study low energy electroweak phenomena at
the level sensitive to the Higgs sector content of the theory.

\medskip

We have proposed a framework for a negative selfconsistency test of the Standard
Model.  It follows from the trivial observation that the models with an effective or
dynamical ultraviolet cutoff are less restrictive than the models with the
physical Higgs particle.  We have suggested experimental conditions that allow
for testing the restrictions following from the interpretation of UV regulator
as a mass of physical particle.  The derived conclusions will be valid for
extensions of the SM that predicts the existence of a massive physical scalar
boson playing the part of the Higgs boson.

 \end{document}